\documentclass[preprints,article,accept,moreauthors,pdftex]{Definitions/mdpi}

\firstpage{1} 
\pubvolume{1}
\issuenum{1}
\articlenumber{0}
\pubyear{2021}
\copyrightyear{2020}
\datereceived{} 
\dateaccepted{} 
\datepublished{} 
\hreflink{https://doi.org/} 

\Title{Project MOMO: Multiwavelength Observations and Modelling of 
OJ 287}

\TitleCitation{Project MOMO: Multiwavelength Observations and Modelling of OJ 287}

\Author{S. Komossa $^{1,*}$, 
D. Grupe $^{2}$, A. Kraus $^{1}$, L.C. Gallo $^{3}$, A. Gonzalez $^{3}$, M.L. Parker $^{4}$, M.J. Valtonen $^{5,6}$, A.R. Hollett $^{3}$, U. Bach $^{1}$, J.L. G\'omez $^7$, I. Myserlis $^{8,1}$ and S. Ciprini $^{9,10}$}

\AuthorNames{S. Komossa, D. Grupe, A. Kraus, et al.}

\AuthorCitation{Komossa, S.; Grupe, D.; Kraus, A.; et al.}

\address{%
$^{1}$ \quad Max-Planck-Institut f\"ur Radioastronomie, 
Auf dem H{\"u}gel 69, 
53121 Bonn, Germany; astrokomossa@gmx.de\\
$^{2}$ \quad Department of Physics, Earth Science, and Space System Engineering, Morehead State University, 235 Martindale Dr, Morehead, KY 40351, USA \\
$^{3}$ \quad  Department of Astronomy and Physics, Saint Mary's University, 923 Robie Street, Halifax, NS, B3H 3C3, Canada \\
$^{4}$ \quad Institute of Astronomy, University of Cambridge, Madingley Road, Cambridge CB3 0HA, UK \\
$^{5}$ \quad Finnish Centre for Astronomy with ESO, University of Turku, FI-20014, Turku, Finland \\
$^{6}$ \quad Department of Physics and Astronomy, University of Turku, FI-20014, Turku, Finland \\ 
$^{7}$ \quad Instituto de Astrofísica de Andalucía-CSIC, Glorieta de la Astronomía s/n, E-18008 Granada, Spain \\
$^{8}$ \quad Instituto de Radioastronomía Milim\'etrica, Avenida Divina Pastora 7, Local 20, 18012, Granada, Spain \\ 
$^{9}$ \quad Istituto Nazionale di Fisica Nucleare (INFN) Sezione di Roma Tor Vergata, Via della Ricerca Scientifica 1, 00133, Roma, Italy \\ 
$^{10}$ \quad ASI Space Science Data Center (SSDC), Via del Politecnico, 00133, Roma, Italy  \\
}

\corres{Correspondence: astrokomossa@gmx.de;  (S.K.)}

\abstract{Our project MOMO (Multiwavelength observations and modelling of OJ 287) consists of dedicated, dense, long-term flux and spectroscopic monitoring and deep follow-up observations of the blazar OJ 287 at $>$13 frequencies from the radio to the X-ray band since late 2015.  In particular, we are using Swift to obtain optical--UV--X-ray spectral energy distributions (SEDs) and the Effelsberg telescope to obtain radio measurements between 2 and 40 GHz. MOMO is the densest long-term monitoring of OJ 287 involving X-rays and broad-band SEDs. 
The theoretical part of the project aims at understanding jet and accretion physics of the blazar central engine in general and the supermassive binary black hole scenario in particular.
Results are presented in a sequence of publications and so far included: detection and detailed analysis of the bright 2016/17 and 2020 outbursts and the long-term light curve; Swift, XMM and NuSTAR spectroscopy of the 2020 outburst around maximum; and interpretation of selected events in the context of the binary black hole scenario of OJ 287 (papers I-IV).  
Here, we provide a description of the project MOMO, a summary of previous results, the latest 
results, and we discuss future prospects.  }


\begin{document}      


\section{Introduction}

 
\subsection{Blazars and supermassive binary black holes} 

Blazars are characterized by their powerful jets of relativistic particles that are launched in the immediate vicinity of the supermassive black holes (SMBHs) at their centers \citep{Blandford2019}. The jets often extend to large distances beyond the host galaxy itself.
The broad-band spectral energy distribution (SED) of blazars exhibits two broad humps of emission \citep{Marscher2009, Ghisellini2015}.
The one at lower energies peaks between the sub-mm and EUV band and sometimes extends into the soft X-ray regime. It is explained as synchrotron radiation from a population of relativistic electrons that form the jet. The second hump peaks in the hard X-ray and/or $\gamma$-ray regime. It is usually explained as inverse Compton (IC) radiation from a population of photons that scatter off the jet electrons. If the synchrotron photons are produced within the jet, their radiation is referred to as  synchrotron-self-Compton radiation (SSC). 
Alternatively, seed photons can be provided by external regions and especially the broad-line region (BLR) or torus (external comptonization; EC).
In addition, or alternatively, hadronic process
may contribute at high energies \citep{Boettcher2019}.  
The accretion disk -- jet interface of blazars  represents one of the most extreme astrophysical environments where high gas densities, strong magnetic fields, and special and general relativistic effects all play a crucial role in shaping the multimessenger emission of these systems.  

Coalescing supermassive binary black holes (SMBBHs), formed in galaxy mergers, are expected to be the loudest sources of low-frequency gravitational waves (GWs) in the Universe \citep{Centrella2010}. They play a key role in galaxy/SMBH formation and evolution scenarios. Therefore, an intense search for wide and close systems in all stages of their evolution is ongoing \citep[][]{Komossa2016}. Wide pairs have been identified by spatially-resolved imaging spectroscopy. However, we rely on indirect methods for detecting the most compact, most evolved systems. These are 
in a  regime where GW emission contributes to orbital shrinkage and they have successfully evolved far beyond the 
`final parsec' in their separation \citep{Begelman1980, Merritt2005, ValtonenKarttunen2006}. Even high-resolution very-long baseline interferometry (VLBI) techniques have so far failed in spatially resolving these systems. Semi-periodicity searches of light curves \citep{Sillanpaa1996, Bon2012, Graham2015, Kelley2019} or of jet-structures \citep{Conway1995, Kun2014, Mohan2016} have therefore been utilized for selecting small-separation SMBBH candidates. 

 Given their well-covered light curves and large-scale jets, blazars are particularly well suited for a search of SMBBHs, and many of the spatially unresolved candidates have been found in blazars \citep{Komossa2016}. 

\begin{figure}
\begin{center}
\includegraphics[clip, width=4.8cm]{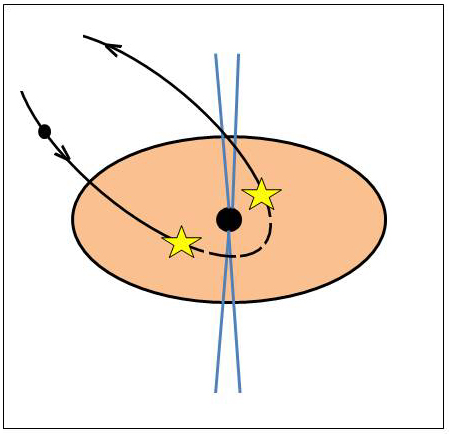}
	\caption{Sketch of the geometry of the binary SMBH model of OJ 287 requiring two disk impacts of the secondary SMBH as it orbits the primary SMBH (not to scale). In addition to these impact flares, the model predicts after-flares when the impact disturbance reaches the inner accretion disk and new jet activity is triggered. 
     }
     \label{fig:binary-orbit}
\end{center}
\end{figure}    
 
\subsection{OJ 287} 

OJ 287 is a nearby, bright blazar with a redshift $z=0.3$ at J2000 coordinates of RA = 08$^{\rm h}$54$^{\rm m}$48.87$^{\rm s}$ and DEC = +20$^\circ$06'30.6". It is  classified as BL Lac object based on its SED and the faintness of its optical broad Balmer lines \citep{SitkoJunkkarinen1985, Nilsson2010}. Its radio--UV SED peaks at low frequencies, leading to a classification of LBL \citep[low-frequency-peaked blazar;][]{Sambruna1996, Padovani1995}, or alternatively as LSP \citep[low synchrotron peak frequency; $<10^{14}$ Hz;][] {Abdo2010} based on another recent SED classification scheme. 
OJ 287 harbours a structured, relativistic radio jet \citep[e.g.][]{Jorstad2005, Agudo2012, Hodgson2017, Britzen2018, Lee2020} and is highly variable in the radio regime in flux \citep{Valtaoja2000, Fuhrmann2014} and polarization \citep{Aller2014, Myserlis2018, Cohen2018}. It has been observed with most of the major X-ray missions and varies strongly in flux and spectrum \citep[e.g.][] {Madejski1988, Urry1996, Donato2005, Ciprini2007, Massaro2008, Seta2009, Marscher2011, K2013, StrohFalcone2013,
Williamson2014, Siejkowski2017, Gallant2018, Komossa2020, Komossa2021a}. At epochs of flaring, OJ 287 is detected with Fermi in the $\gamma$-ray band \citep{Abdo2009, Agudo2011, Hodgson2017}. 

OJ 287 is among the best candidates to date for hosting a compact SMBBH \citep{Sillanpaa1988, Valtonen2008, Dey2018, Valtonen2019, Laine2020}. 
Its unique optical light curve dates back to the 1880s and shows characteristic double-peaks every $\sim$12 years \citep{Pursimo2000, Villforth2010, Hudec2013, Dey2018}
that have been interpreted as arising
from the orbital motion of a SMBBH, with an orbital period on that order ($\sim$9 yrs in the system's rest frame). 
Following the discovery that major optical outbursts of OJ 287 repeat \citep{Sillanpaa1996}, different variants of binary scenarios of OJ 287 have been considered. 
The best explored model explains the double peaks as
events when the secondary SMBH intercepts the disk around the primary SMBH twice during its orbit (`impact flares' hereafter \citep{Lehto1996, Valtonen2019}; our Fig. \ref{fig:binary-orbit}). 
This model requires a compact SMBBH with a primary SMBH of mass $1.8\times10^{10}$ M$_{\odot}$ and spin 0.38, and a secondary of mass $1.5\times10^8$ M$_{\odot}$ with a semi-major axis of 9300 AU on an eccentric, tilted orbit.  The secondary's orbit is subject to general-relativistic (GR) forward precession, causing a change in the time interval of subsequent impact flares.  
These are not always separated by 12 yr. Their separation varies with time and in a predictable manner. 
According to hydrodynamic simulations \citep{Ivanov1998}, the secondary's impact on the disk drives a two-sided supersonic bubble of hot, optically thick gas that expands and cools. 
Once the gas becomes optically thin, it starts emitting and at that time the flare becomes observable.
The last two impact flares were reported in 2015 and 2019 \citep{Valtonen2016, Laine2020}. 
The one in 2019 was not observable with Swift or from the ground (in the optical band) due to the proximity of OJ 287 to the sun. The Spitzer observatory was used instead. 
At epochs of impact flares, there is an additional optical-IR emission component that may extend into the EUV or soft X-rays with a bremsstrahlung temperature of the order 10$^{5-6}$ K {\footnote{e.g. $T_{\rm brems}=3^{+6}_{-2} \times 10^5$\,K during the 2005 flare; \citep{Valtonen2012}}} \citep{Ciprini2007, Valtonen2012, Valtonen2019}, and the polarization of the {\em {total}} optical flux decreases \citep{Smith1987, Valtonen2008, Valtonen2016}.
In addition to the impact flares, the model predicts `after-flares' when the impact disturbance reaches the inner accretion disk \citep{Sundelius1997, Valtonen2009} and new jet activity is triggered. 

Recent simulations of major mergers (SMBH mass ratios $q > 0.1$) typically predict a scenario in which a {\em{circum-binary}} disk forms \citep{Liu2004, MacFadyen2008, Cuadra2009}. 
Simulations for a wider range of mass ratios $q<<1$ have been carried out in 2D assuming that the secondary is in the disk plane \citep[e.g.][]{Haiman2009, D'Orazio2016, Duffell2020}. 
These models therefore assume a different geometry than required for the OJ 287 binary where the secondary's orbit is highly inclined and the mass ratio is $q < 0.01$. We therefore expect a different evolutionary path of the binary system with a disk bound to the primary.

When we mention the binary model of OJ 287 then we refer to the model developed by Valtonen and collaborators \citep[e.g.][]{Sillanpaa1988, Sillanpaa1996, Valtonen1996, Lehto1996, Valtonen2008, Dey2018, Valtonen2019, Laine2020} unless noticed otherwise. Other variants of binary scenarios were considered in the literature as well \citep[e.g.][]{Katz1997, Villata1998, Valtaoja2000, Liu2002}, explaining aspects of the light curve of OJ 287 with a non-precessing binary impacting the disk only once,
or with Doppler-boosting of a jet sweeping our line-of-sight
either due to Newtonian precession or 
due to the orbital motion of a jet-emitting secondary SMBH.
The alternative of a precessing accretion disk was explored as well \citep{Britzen2018, Liska2018}. Such a scenario has to address the cause of disk precession (that could again be due to a binary). 
%
However, none of these scenarios has so far seen the detailed modelling and predictions over many years that are unique to the binary variant of OJ 287 first proposed by Valtonen et al. \citep{Valtonen1996, Lehto1996}.   

\begin{figure}
\begin{center}
\includegraphics[clip, trim=1.8cm 2.1cm 2.2cm 0.3cm, angle=-90, width=11cm]{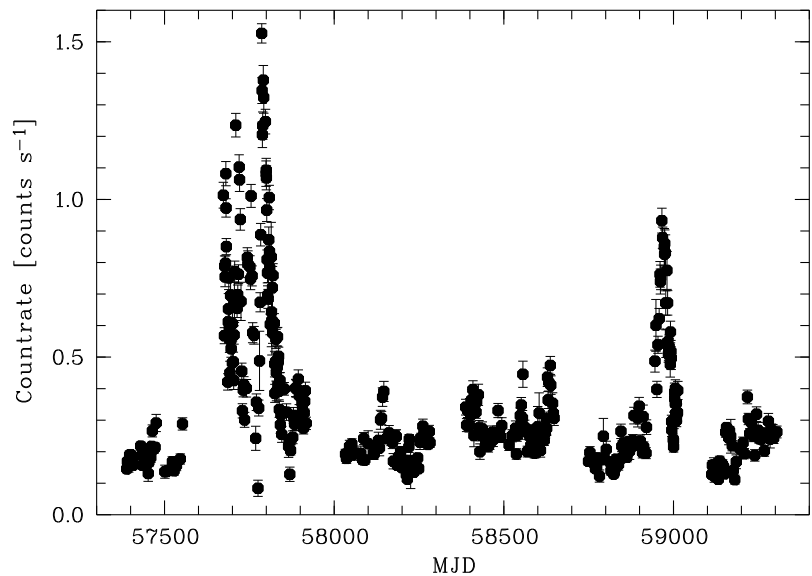}
	\caption{Swift (0.3-10 keV) X-ray light curve of OJ 287 between 2016 January and 2021 March 31. Two prominent outbursts occurred in 2016/17 and 2020. Error bars are always plotted but are often smaller than the symbol size. The majority of data was obtained in the course of the MOMO project. 
     }
     \label{fig:Swift_Xlight-MOMO}
\end{center}
\end{figure}     
 
\section{MOMO project description}

\subsection{Motivation, set up, and key goals}

Among the blazar population, OJ 287 stands out in several respects: First, it is among the best candidates to date for hosting a compact SMBBH. 
The timing of the impact flares and after-flares in the best-explored and best-tested binary model of OJ 287 depends crucially on GR precession effects of the binary orbit, warps in the accretion disk,
as well as the binary orbital shrinkage due to GW emission \citep{Valtonen2008, Valtonen2016, Dey2018, Laine2020}. OJ 287 therefore is an excellent candidate for studying strong-field relativistic effects and for searching for consequences of GW-inspiral well before the launch of space-based GW interferometers and well before more sensitive pulsar timing arrays
start operating in the framework of SKA-II \citep{Yardley2010}. 
Second, independent of its possible binary nature, OJ 287 is one of the very few LBLs (low-energy peaked blazars) which have their synchrotron component(s) extending into the soft-X-ray regime, optimally constraining the high-energy end of the electron population most sensitive to heating/cooling processes, while at the same time the inverse Compton component is detected in the same (XMM-Newton and Swift) X-ray band \citep{Komossa2021a}. Third, OJ 287 is currently undergoing a phase of exceptional activity,
and has been known in the past for the great brightness of some of its optical outbursts, reaching 12th mag and making it one of the brightest members of the blazar population.   
For these reasons, we have initiated the program MOMO (Multiwavelength Observations and Modelling of OJ 287; \citep[e.g.][]{Komossa2017,Komossa2021a}) which started in late 2015 after the report of a bright optical flare in December 2015 interpreted as impact flare in the binary SMBH model \citep{Valtonen2016}.
The goal is to cover at least one $\sim$12 yr cycle of OJ 287 (Fig. \ref{fig:Swift_Xlight-MOMO}). 

MOMO rests on an observational and a theoretical pillar. 
The observational part consists of multi-year, multi-frequency flux and spectroscopic monitoring at $>$13 frequencies from the radio to the X-ray regime.
The Neil Gehrels Swift observatory \citep[Swift hereafter;][]{Gehrels2004} and the Effelsberg radio telescope play a central role. 
In addition, deeper follow-up observations at other facilities are triggered in case of outbursts, deep low-states or other exceptional states of special interest. MOMO is the densest long-term monitoring of OJ 287 involving X-rays, the UV, 
and broad-band SEDs with a cadence as short as 1 d. 
On the theoretical side, focus is on SED modelling, on the jet physics that drive the cross-band delays and the characteristic timescales obtained from discrete correlation function (DCF) and power spectral density analyses, on the disk-jet connection, on the distinction between different jet precession and binary scenarios that make distinctly different predictions for the timing and SEDs of flares, and on aspects of the recent evolution of the binary system. The upcoming Sections 2.2--2.4 will provide more details of the 
model constraints we can extract in each waveband. 

A few individual Swift (and XMM-Newton) observations of OJ 287 are timed with the Event Horizon Telescope (EHT; \citep{EHT2019}) to obtain quasi-simultaneous SEDs \citep{Komossa2021a}. Independent of the MOMO program, OJ 287 is a prime target of the EHT and other VLBI observations. 

All data obtained by us are analyzed quickly. 
The community is alerted rapidly in {\sl{Astronomer's Telegrams}} about outbursts or other noteworthy states of OJ 287 we detect with Swift or at Effelsberg
(ATel \#8411, \#9629, \#10043, \#12086, \#13658, \#13702, \#13785, \#14052) to enable them to take additional multiwavelength data not covered by our own project. 

\begin{figure}
\centering
\includegraphics[clip, trim=1.8cm 5.6cm 1.3cm 2.6cm, width=\columnwidth]{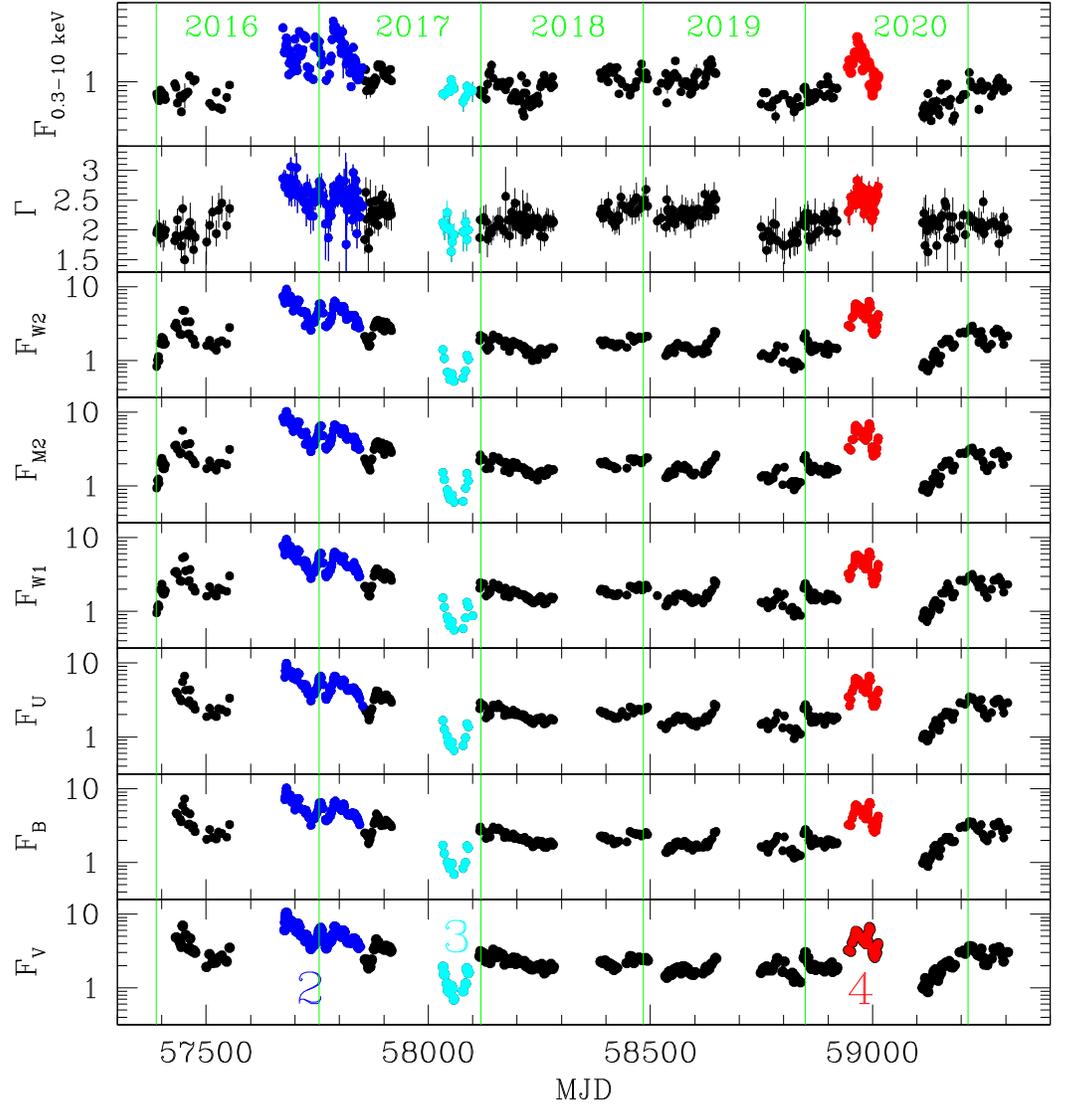}
\caption{Swift light curve of OJ 287 between 2016 January and 2021 \citep{Komossa2017, Komossa2020, Komossa2021a} now expanded until 2021 March 31. The majority of data was obtained by us in the course of the MOMO project. 
The X-ray flux (in the observed 0.3-10 keV band, and corrected for Galactic absorption) and the UV--optical fluxes (corrected for Galactic extinction) are given in 10$^{-11}$ erg s$^{-1}$ cm$^{-2}$. $\Gamma_{\rm x}$ is the X-ray power-law photon index.
January 1st of each year between 2016 and 2021 is indicated by a green vertical line. Error bars are always plotted but are often smaller than the symbol size. Three epochs are marked in colour: the 2016/17 outburst (dark blue), the 2017 UV--optical deep fade (light blue), and the April--June 2020 outburst (red).  
}   
    \label{fig:lc-Swift-fluxes2016}
\end{figure}

\subsection{MOMO-X}

We are using the Swift X-ray telescope (XRT; Burrows2005) to obtain  X-ray spectra of OJ 287 
along with occasional deeper spectroscopic observations with XMM-Newton and NuSTAR \citep{Komossa2020}. The Swift XRT operates between 0.3--10 keV. The majority of exposure times is in the range 1--2 ks. 
All Swift X-ray spectra are fit with absorbed, single-component power laws of photon index $\Gamma_{\rm x}$. The absorption is fixed at the Galactic foreground absorption toward OJ 287 (Hydrogen column density $N_{\rm H,Gal} = 2.49 \times 10^{20}$ cm$^{-2}$ \citep{Kalberla2005}; see \citep{Komossa2020} for further details of the Swift data analysis). 
The typical cadence of observations ranges between 1--5 d and is higher at epochs of rapid flux changes (outbursts, deep minima). Each year, our Swift observations of OJ 287 are interrupted by $\sim$ 3 months because OJ 287 is unobservable due to its proximity to the sun. 

The X-ray band traces the synchrotron emission and/or inverse-Compton emission, but is also sensitive to accretion-disk emission. Given the large mass of the primary SMBH required by the binary model, we do not expect any significant extension of any thermal (multi-temperature black body) accretion-disk emission into the soft X-ray regime \citep{Done2012}. However, we do expect the presence of an accretion-disk corona in X-rays, even though at most epochs it will be undetectable since it is much fainter than the jet emission (as commonly the case in BL Lac objects).    

\subsection{MOMO-UO}

We are also using the Swift UV-optical telescope \citep[UVOT;][]{Roming2005} to obtain dense coverage in all 3 optical and all 3 UV filters (Tab. \ref{tab:overview}).{\footnote{Given the redshift of OJ 287, UVOT filter central wavelengths \citep{Poole2008} correspond to $\lambda_{0}=1476$\AA ~(UV-W2) and $\lambda_{0}=4187$\AA ~(V), respectively, in the restframe.}}
Occasional follow-up optical spectroscopy at ground-based telescopes is obtained to search for spectroscopic (emission-line) activity at select epochs of outbursts or low-states.   
The UVOT data are taken quasi-simultaneous with the XRT data, and at the same cadence. Exposure times for the UVOT are on the order 1--2 ks, where under normal circumstances the filters V:B:U:W1:M2:W2 are observed with a ratio of 1:1:1:2:3:4 of the total exposure time, respectively \citep{Grupe2010}. UV--optical fluxes are corrected for Galactic extinction \citep{Schlegel1998}. 

\begin{figure}
\begin{center}
\includegraphics[clip, trim=1.0cm 0.0cm 1.4cm 2.1cm, width=9.5cm]{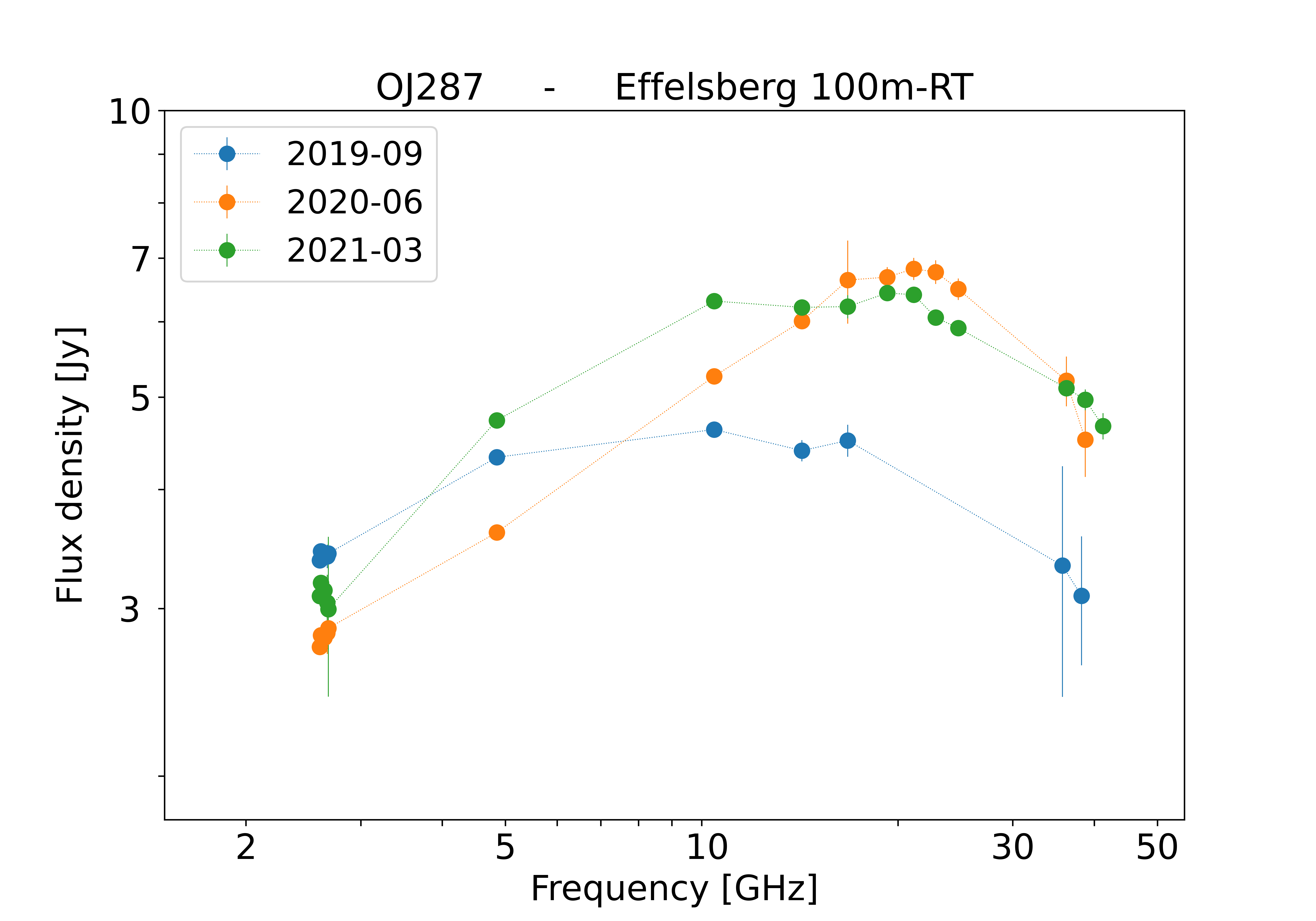}
	\caption{Representative recent radio SEDs of OJ 287 between 2.6 and 40 GHz obtained at the Effelsberg 100m telescope in the course of the MOMO project (taken from ref. \citep{Komossa2021b} with permission of the editors). 
     }
     \label{fig:EB_radio_spectra}
\end{center}
\end{figure}

While the optical-UV emission of OJ 287 is dominated at most times by synchrotron radiation, the dense SED coverage allows us to search for temporary accretion-disk or other thermal emission contributions as well; rarely expected as OJ 287 is a BL Lac, except for epochs of thermal emission predicted by the binary SMBH scenario (Sect. 1.2) or at epochs when the jet emission is in a very deep low-state such that the faint, long-lived accretion-disk emission becomes detectable{\footnote{Note that OJ 287 has a BLR detected at continuum low-states, independently implying the presence of a significant but faint accretion disk \citep{SitkoJunkkarinen1985, Nilsson2010}.}}. 

The Swift UVOT and XRT light curve of OJ 287 between 2016 and 2021 is shown in Fig. \ref{fig:lc-Swift-fluxes2016} (see \citep{Komossa2021a, Komossa2021c} for the complete Swift light curve starting in 2005). UV--optical fluxes are displayed as flux densities multiplied by filter central frequencies. Most of the data were obtained by us in the course of the MOMO project, but we have added public archival data as well. 

\subsection{MOMO-Radio}
In the radio regime, we are using the Effelsberg 100m radio telescope to obtain multi-frequency flux and spectral measurements at dense cadence of typically twice per month (PI: S. Komossa; program identifications 99-15, 19-16, 12-17, 13-18, 75-19, and 65-20) since late 2015 \citep{Komossa2015}. Frequencies between 2.6 and 40 GHz are employed (Fig. \ref{fig:EB_radio_spectra}, Tab. \ref{tab:overview}). Data reduction and analysis procedures follow ref. \citep{Kraus2003, Myserlis2018}. 
In the radio regime, a coverage of OJ 287 is possible even at those epochs each year where OJ 287 is unobservable with ground-based optical telescopes and with Swift due to its solar proximity. The Effelsberg telescope can still observe sources at distances larger than 1 degree from the sun. First radio results \citep{Komossa2015, Myserlis2018} obtained in the course of the MOMO project revealed strong changes in the polarization and EVPA of OJ 287.  

\begin{figure}
\begin{center}
\includegraphics[clip, trim=1.8cm 2.3cm 1.0cm 0.3cm, angle=-90, width=9cm]{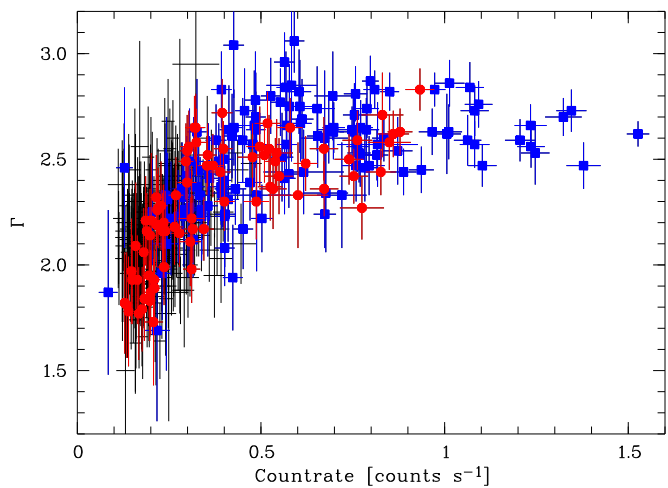}
	\caption{Softer-when-brighter variability pattern of OJ 287 since January 2016 based on MOMO and first reported by ref. \citep{Komossa2017}. Outburst epochs are marked in colour (2016/2017: blue, 2020: red). 
     }
     \label{fig:softer-when-brighter}
\end{center}
\end{figure}

Radio emission traces the synchrotron components.
Ongoing measuremenst are used to: (1) Obtain the flux density and its evolution, especially  in the  time intervals around the binary impact flares and after-flares. (2) Time the radio high-state(s), especially with respect to the optical maxima and our own multi-wavelength data, but also including $\gamma$-rays. 
(3) Use the radio and cross-correlated optical, UV, X-ray, and other MWL data, and spectral turnover frequency, to constrain the jet physics and magnetic field, including with low- and high-level flaring that goes on independent of the binary's presence.  
(4) Distinguish between different SMBBH scenarios and explore facets of the best-developed binary model, based on distinctly different predictions in the radio regime for the first and second optical peak of the double peaks and for the possible after-flares.   
For instance, if major optical flares 
are thermal in nature, they will not be accompanied by radio flares{\footnote{except for epochs, where the secondary SMBH may be temporarily accreting and launching a short-lived jet; \citep{Pihajoki2013, Dey2021}}}. If both optical peaks are synchrotron peaks, we expect two radio flares, with polarization evolution following synchrotron theory.  

\begin{specialtable}[H] 
\caption{Summary of the instruments and wavebands that form the core of MOMO. (Note that we only list the radio frequencies most recently employed. Some receivers have changed in the past, and at selected epochs a larger range of frequencies was used.) \label{tab-overview}}
\begin{tabular}{llc}
\toprule
\textbf{mission}	& \textbf{instrument}	& \textbf{waveband or central wavelength}  \\
\midrule
Swift & XRT & 0.3--10 keV  \\
Swift & UVOT--W2 & 1928\AA \\ 
      & UVOT--M2 & 2246\AA \\ 
      & UVOT--W1 & 2600\AA \\ 
      & UVOT--U & 3465\AA \\ 
      & UVOT--B & 4392\AA \\ 
      & UVOT--V & 5468\AA \\ 
Effelsberg 100m & S7mm  & 33--50 GHz \\ 
                & S14mm  & 18--26 GHz \\ 
                & S20mm  & 12--18 GHz \\ 
                & S28mm  & 10.45 GHz \\ 
                & S60mm  &  4.85 GHz \\ 
                & S110mm &  2.55 GHz \\ 
\bottomrule
\label{tab:overview}
\end{tabular}
\end{specialtable}


\section{MOMO results}
Here, we give an overview of results obtained so far (papers I-IV), we use the most recent results to update light curves and correlation diagrams, and we provide new results and analyses using the most recent Swift observations and Effelsberg multi-frequency radio observations since 2019. This is an ongoing project and more results will be reported in the future. 

\subsection{Long-term light curve} 

The Swift long-term X-ray--UV--optical light curve of OJ 287 (Fig. \ref{fig:Swift_Xlight-MOMO, fig:lc-Swift-fluxes2016}) is characterized by high-amplitude variability. The flux is rarely constant over extended periods of time. Two bright, super-soft X-ray outbursts were discovered in 2016/17 \cite{Komossa2017} and 2020 \citep{Komossa2020}.  The optical--UV and X-rays are closely correlated.  In addition, the light curve reveals multiple episodes of mini-flaring on the timescale of weeks to months between 2016 and 2021. 
Overall, the level of activity of OJ 287 is significantly higher and the X-ray spectra are steeper than during the early Swift observations of OJ 287 in 2005--2007 (\citep{Massaro2008}; see \citep{Komossa2021a, Komossa2021c} for the long-term Swift light curve since 2005) when X-ray countrates dropped below 0.1 cts/s and the photon index was as flat as  $\Gamma_{\rm x} \approx 1.5$.  

The recent Effelsberg light curve 
reveals a radio flare that accompanies the 2020 outburst.  A second flare of comparable peak flux is also seen in 2021, when the X-rays remained in a rather low state and the optical-UV showed mini-flares superposed on a broader multi-months flare with intermediate peak flux, not reaching the levels of the 2016/17 and 2020 outbursts. 

\begin{figure}
\begin{center}
\includegraphics[clip, width=7.5cm]{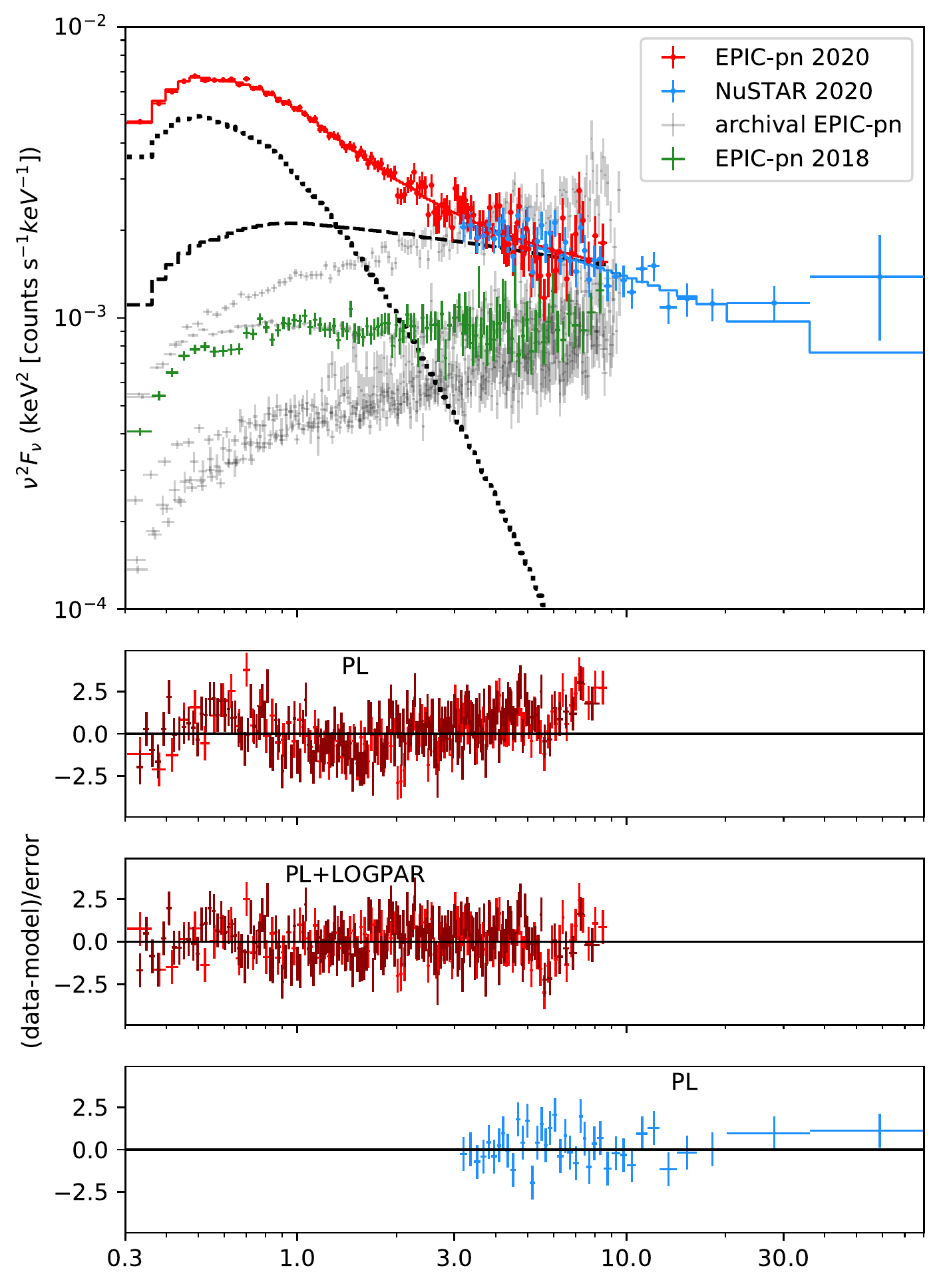}
	\caption{XMM-Newton EPIC-pn (red) and NuSTAR (blue) observations we obtained during the 2020 outburst of OJ 287 \citep{Komossa2020}. Note the giant soft X-ray excess detected with XMM-Newton. Our 2018 XMM-Newton observation (taken quasi-simultaneous with the EHT; \citep{Komossa2021a}) is shown in green, and all previous PI and archival XMM-Newton spectra between 2005 and 2015 are displayed in grey for comparison. The three lower panels show the residuals of spectral fits between 0.3-60 keV from a  single power law and a power law plus logarithmic parabolic power law fit to the XMM-Newton data, and a power law fit to the NuSTAR data, respectively.
     }
     \label{fig:XMM-Nustar2020}
\end{center}
\end{figure}

\subsection{X-ray spectroscopy}

Swift monitoring observations are typically of 1--2 ks duration, and therefore do not allow multi-component spectral fits. Single power-law fits are well-constrained and provide important information on the dominant emission mechanism; super-soft synchrotron emission during epochs of outbursts as steep as $\Gamma_{\rm x} \simeq 3$ and flat inverse-Compton emission  with  $\Gamma_{\rm x} \simeq 1.5$. A strong softer-when-brighter variability pattern is seen (Fig. {\ref{fig:softer-when-brighter}), explained as increasing dominance of the soft synchrotron component as the flux increases. 

XMM-Newton spectra allowed us to perform spectral decompositions into two emission components during each observation \citep{Komossa2021a}. In particular, the spectrum we obtained during the 2020 outburst is characterized by super-soft synchrotron emission well described by a logarithmic parabolic power-law model and a flatter emission component, detected in our NuSTAR spectrum up to $\sim$70 keV (\citep{Komossa2020}; our Fig. \ref{fig:XMM-Nustar2020}). 
The break energy, where the soft and hard spectral components intersect, is located near 1--2 keV in most XMM-Newton spectra of OJ 287.

In summary, Swift and XMM-Newton results have established OJ 287 as one of the most spectrally variable blazars known in the X-ray band with spectral indices (using the power-law description) between $\Gamma_{\rm x}$= 1.5 ... 3.

\begin{figure}
\begin{center}
\includegraphics[clip, trim=1.6cm 1.2cm 1.9cm 0.3cm, angle=-90, width=0.49\linewidth]{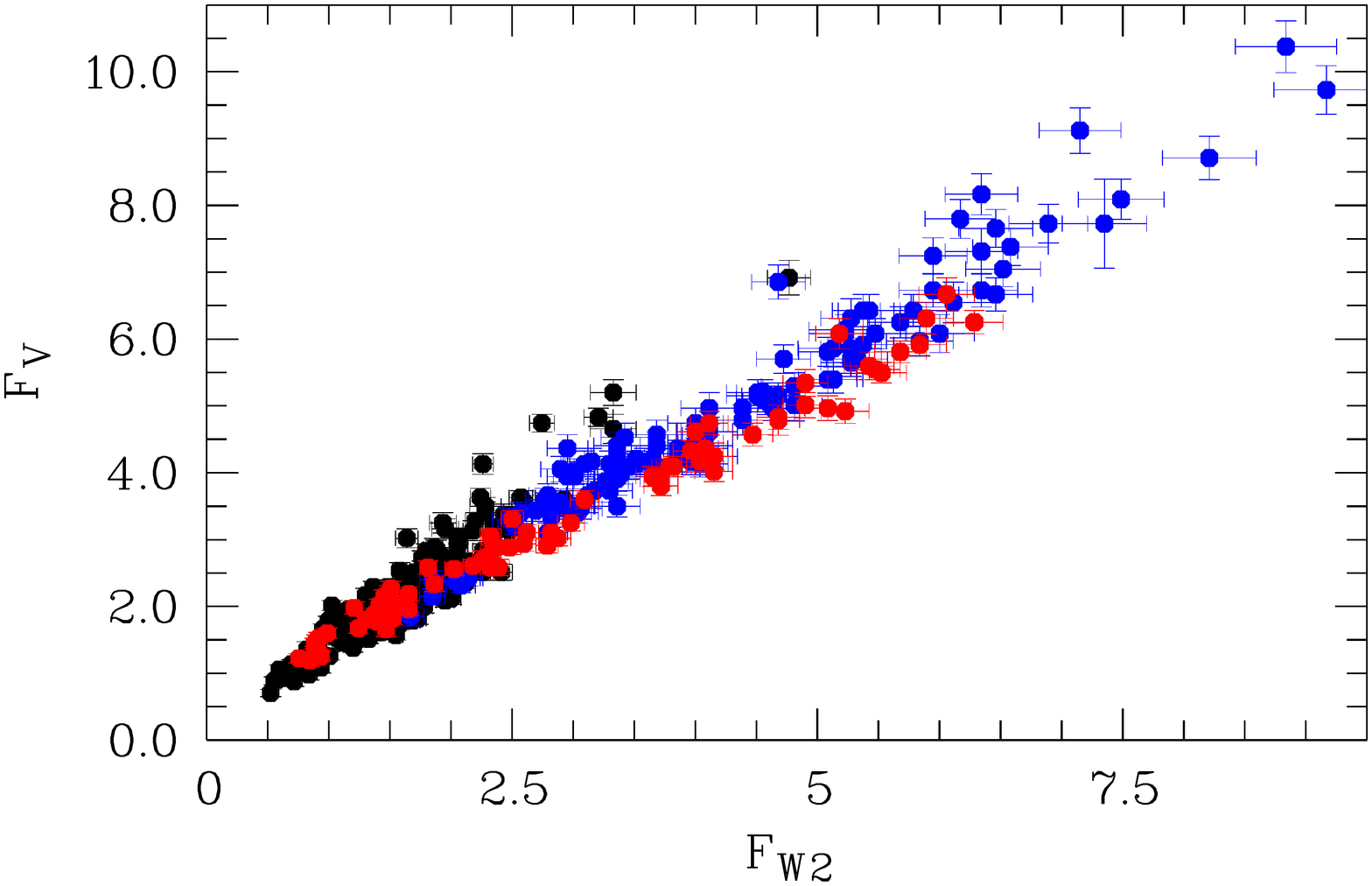}
\includegraphics[clip, trim=1.6cm 1.3cm 1.9cm 0.3cm, angle=-90, width=0.49\linewidth]{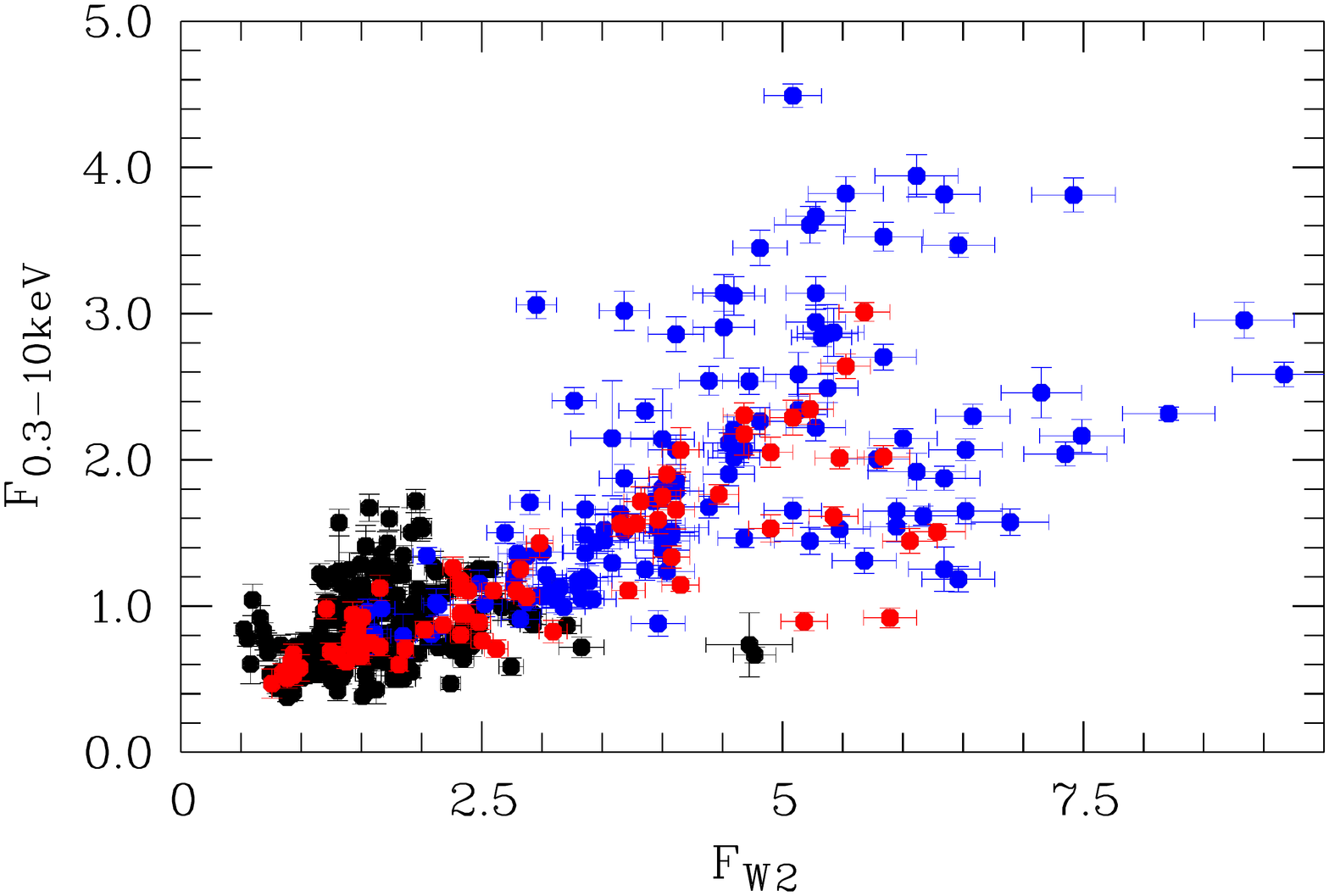}
	\caption{Optical(V)--UV(W2) flux correlation (left) and X-ray-UV(W2) flux correlation (right) between 2016 January and 2021 March 31. Epochs including the outbursts of 2016/17 (blue) and 2020 (red) are marked in colour. Fluxes are reported in units of 10$^{-11}$ erg cm$^{-2}$ s$^{-1}$.
     }
     \label{fig:UV-X-corr}     
\end{center}
\end{figure}

\subsection{Inter-band correlations, DCF}

Characterization of variability timescales and cross-band correlations provide constraints on emission processes and particle distributions \citep{Sokolov2004, Kirk1998, Marscher2014} and form an important part of MOMO. 
%

The X-ray band exhibits a strong softer-when-brighter variability pattern (Fig. \ref{fig:softer-when-brighter}) implying an increasing contribution of a synchrotron component that completely dominates the high-state spectra. 
While the UV and optical fluxes are closely correlated at all times, the X-rays correlate with the optical--UV but show larger scatter (Fig. \ref{fig:UV-X-corr}). 
To analyze cross-band correlations in more detail, we have carried out a discrete correlation function (DCF; \citep{Edelson2017}) analysis for 5 different epochs between 2015 and 2020 \citep{Komossa2021c}. Epoch 1 (2015) was added from the Swift archive (PI: R. Edelson) because of the dense cadence of twice/day for a time interval of $\sim$6 weeks. Epochs 2 (2016/17), 3 (2017/18), 4 (2018/19) and 5 (2019/20) are from the MOMO program. These epochs are of 9 months duration each, and are separated from each other by a 3 month gap where OJ 287 is unobservable with Swift each year. Epochs 2 and 5 include the outbursts, epochs 1, 3 and 4 represent low-level activity.  
The DCF analysis has shown that the optical and UV are always closely correlated with a lag time on the order of 0 d ($\tau = 0\pm1$ d during the epoch in 2015 where cadence was highest).  Instead, X-rays only show zero lag w.r.t the optical--UV at outburst states. At low-states, the X-rays are leading or lagging, by $\tau = -18 ... +7$ d (Fig. \ref{fig:DCF}). 

\begin{figure}
\includegraphics[clip, width=\columnwidth]{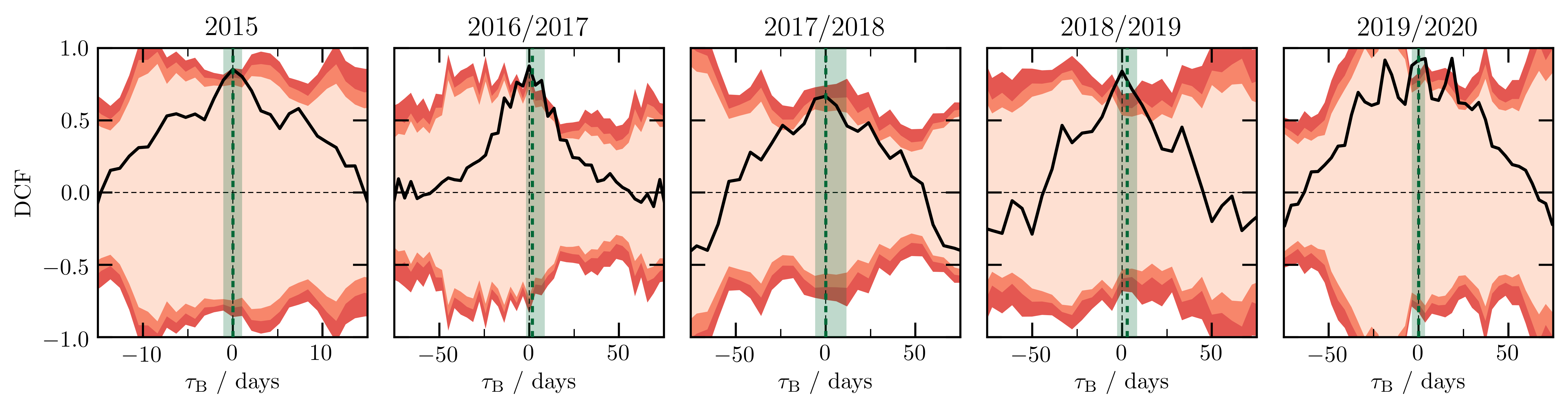}
\includegraphics[clip, width=\columnwidth]{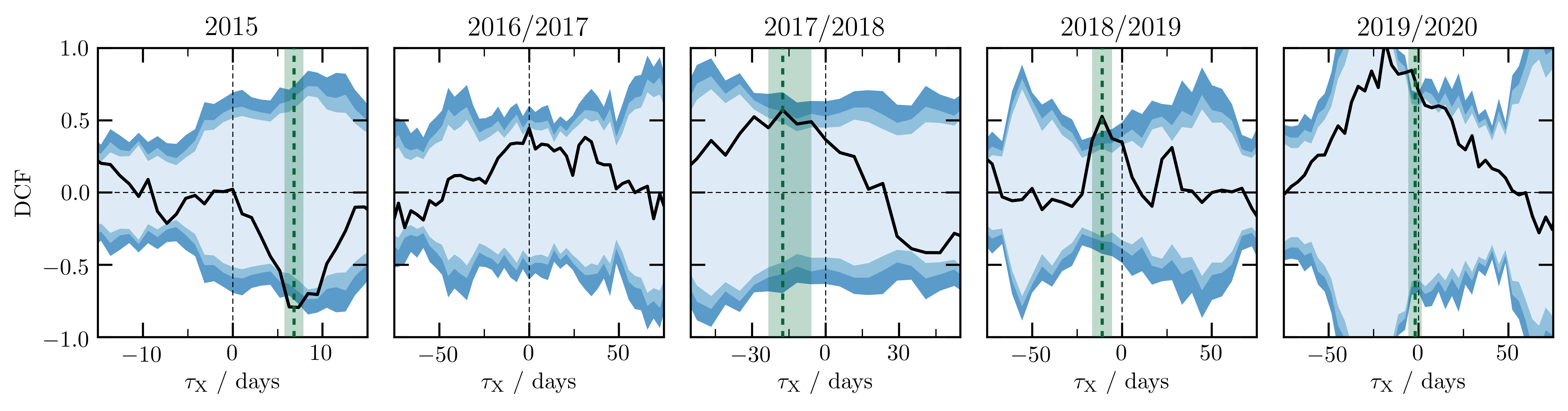}
\caption{Discrete correlation functions [upper panel: optical(B)--UV(W2) flux, lower panel: X-ray--UV(W2) flux] computed for five epochs between 2015 and 2020 shown from left to right (epochs 1, 3, and 4 represent states of low-level activity; epochs 2 and 5 are dominated by outbursts).
Filled regions indicate the $\pm90^{\mathrm{th}}$ 
(light colour), $\pm95^{\mathrm{th}}$ (medium colour), and $\pm99^{\mathrm{th}}$ (dark colour) percentiles from $N=10^3$ light-curve simulations. 
Horizontal dashed lines indicate $\mathrm{DCF}=0$,
and vertical dashed lines indicate $\tau_{\mathrm{B,X}} = 0~\mathrm{days}$. The vertical green line marks the measured lag and its error. 
Negative $\tau_{\mathrm{X}}$ values indicate X-rays leading W2, positive values indicate lagging. 
During the epoch 2017/2018, the UV--optical deep fade revealed an independent X-ray component not following the deep fade. Therefore, the X-ray--UV DCF was only carried out after MJD 58100 
for this epoch \citep{Komossa2021c}.
UV and optical fluxes are always closely correlated with a lag consistent with zero days.  X-rays closely follow the optical--UV during outburst epochs (2016/2017 and 2019/2020), but are lagging (2015) or leading (2017/2018 and 2018/2019) during epochs of low-level activity. 
     }
     \label{fig:DCF}
\end{figure}

The near-zero lag between all bands during outbursts is consistent with synchrotron theory \citep{Kirk1998}. 
During epochs of low-level activity, the X-rays are no longer dominated by soft synchrotron emission. The X-ray spectra are much flatter, and indicate an inverse Compton contribution. Both, leads and lags on the timescale of days to weeks between the optical and the high-energy emission are predicted by synchrotron-self-Compton (SSC) and  external comptonization (EC) models \citep{Sokolov2004, Sikora2009}. SSC models predict shorter timescales than we observe. We therefore favor EC. The BLR (detected at continuum low-states; \citep{Nilsson2010}) is a plausible source of external seed photons.  
The correlated variability we detect between the  optical-UV and the X-rays (those not of synchrotron origin) at low-states favors a leptonic jet model of OJ 287 since these two parts of the SED are expected to vary more independently in hadronic models.

\subsection{SEDs}

\begin{figure}
\begin{center}
\includegraphics[clip, width=11cm]{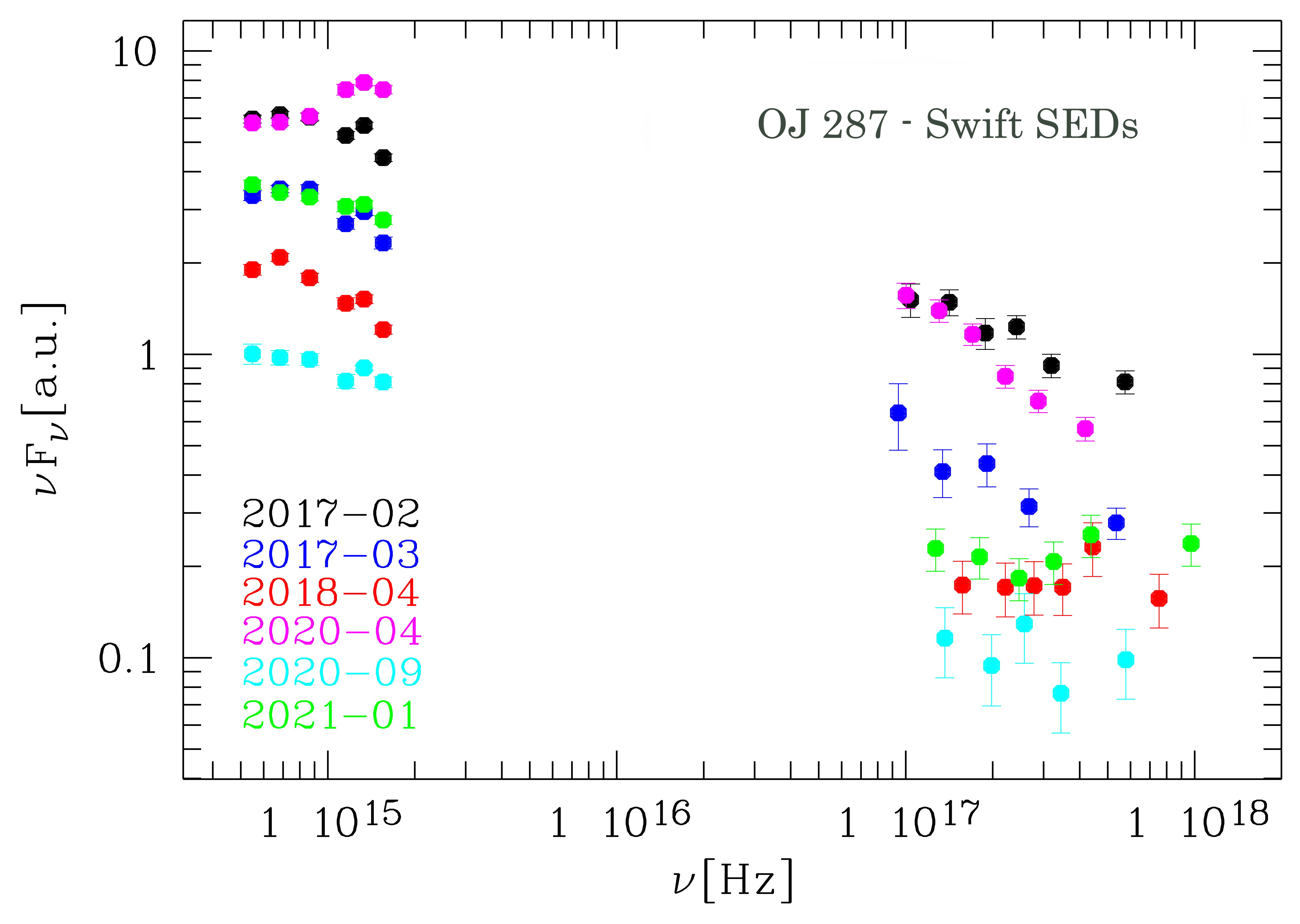}
\caption{Observed SEDs of OJ 287 obtained with Swift at selected epochs: the 2016/17 and 2020 outbursts at peak (February 2017 and April 2020), March 2017, April 2018,  
the September 2020 low-state \citep{Komossa2021a}, and an epoch from January 2021.  The X-ray spectral steepening at high-states is due to the increasing contribution of the synchrotron component(s).  
}
\label{fig:Swift_SEDs}
\end{center}
\end{figure}

\begin{figure}
\begin{center}
\includegraphics[clip, trim=0.8cm 1.5cm 0.8cm 0.0cm, width=9.0cm, angle=-90]{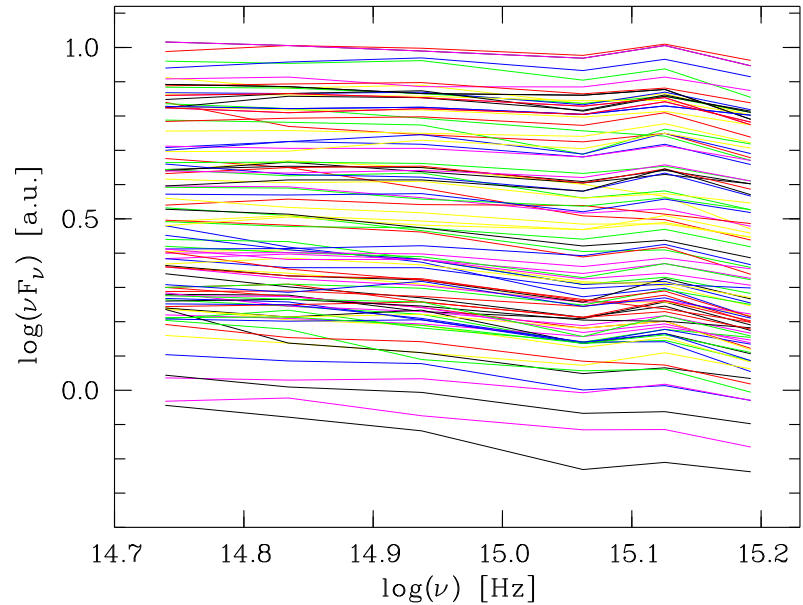}
\caption{Zoom on selected Swift UVOT SEDs  of OJ 287 since 2016. 
}
\label{fig:Swift_SEDs_UVOT}
\end{center}
\end{figure}

One of the long-term goals of MOMO is modelling of the SEDs of OJ 287 at different flux and spectral states. Selected broad-band Swift SEDs are shown in Fig. \ref{fig:Swift_SEDs}. 
The X-ray and UVOT spectral shapes are correlated, with softer X-ray spectra corresponding to bluer optical-UV spectra, interpreted as an increasing dominance of synchrotron component(s) at high-states.

\subsection{Individual outstanding epochs and comparison with  predictions of the SMBBH model}

Here we comment on the most noteworthy events in the light curve of OJ 287 since 2016 (see Tab. \ref{tab:tab2} for a summary).

\begin{specialtable}[H] 
\caption{Summary of outstanding events in the Swift light curve or OJ 287 and/or predictions from the binary scenario between 2016 and 2021.\label{tab2}}
\begin{tabular}{llll}
\toprule
\textbf{event}	& \textbf{obs. date}	& \textbf{waveband} & \textbf{comments} \\
\midrule
outburst & Sep 2016--Apr 2017 & all & consistent with binary after-flare \citep{Valtonen2017}; \\
         &                    &     &   supersoft X-ray spectrum, non-thermal  \\                 
deep fade & Oct--Dec 2017 & UV, opt & symmetric fade, not seen in X-rays \\
 &&& \\
impact flare & July 2019 & IR & predicted; detected with Spitzer; \citep{Laine2020} \\
 &&& \\
outburst & Apr--Jun 2020 & all & consistent with binary after-flare \citep{Komossa2020}; \\
         &                    &     &   supersoft X-ray spectrum,
non-thermal\\
\\ low-state & Sept 2020 & UV, opt & \\
precursor-flare & Dec 2020 & opt & predicted; \citep{Pihajoki2013} \\ 
\bottomrule
\label{tab:tab2}
\end{tabular}
\end{specialtable}

\subsubsection{2016/17 outburst}

OJ 287 underwent a bright outburst starting in September 2016 and extending into early 2017 \citep{Komossa2017, Komossa2020}. This was the brightest X-ray outburst of OJ 287 recorded with Swift. 
Despite early speculations about an accretion 
flare at this epoch as a possible explanation for the softness of the observed X-ray spectrum, multiple arguments then clearly established this outburst as
non-thermal in nature: 
First, with Swift we detect X-ray
flux doubling timescales as short as 4 days; shorter than the light-crossing time at the last stable orbit of the accretion disk around the primary SMBH ($M_{\rm SMBH} = 1.8 \times 10^{10}$ M$_{\odot}$)
ruling out a primary's disk origin. 
Second, our optical--UV DCF results are consistent with synchrotron theory, but the lags are too small for accretion-disk reverberation of a SMBH with mass as low as $\sim$10$^{8}$ M$_{\odot}$. 
Third, the Swift X-ray spectra are well explained by a soft synchrotron emission component and show the very same softer-when-brighter variability pattern also seen during the non-thermal 2020 synchrotron outburst (\citep{Komossa2020}; our Fig. \ref{fig:softer-when-brighter}).
Further, the outburst was accompanied by a radio
flare \citep{Myserlis2018, Lee2020} and by VHE emission detected by VERITAS 
\citep{OBrien2017}, and the optical band showed high levels of polarization \citep{Valtonen2017} as did the radio \citep{Goddi2021}. The 2016 outburst, independently detected during ground-based optical monitoring, was interpreted as an after-flare predicted by the binary SMBH model \citep{Valtonen2017}. 

\subsubsection{2017 UV--optical deep fade}


During October to December 2017, the Swift light curve of OJ 287 exhibits a remarkable, sharp, symmetric deep fade and recovery in the UV and optical band \citep{Komossa2020, Komossa2021c}, at first glance reminiscent of an occultation event (the deep fade is marked in light blue as feature number 3 in the long-term light curve; Fig. \ref{fig:lc-Swift-fluxes2016}){\footnote{The deep fade was independently noticed in ground-based optical monitoring and was used to obtain imaging of the host galaxy of OJ 287 while the blazar glare itself was least affecting the host detection \citep{Nilsson2020}.}}. However, we can rule out the passage of a dusty cloud along our line-of-sight, because the expected optical--UV reddening is not observed. The optical/UV flux ratio remains constant during the event. Further, it is unlikely that the secondary SMBH temporarily deflected the jet between primary and observer \citep{Takalo1990} since the secondary was expected to be behind the accretion disk during the event \citep{Dey2018, Dey2021}. 
The UV--optical deep fade reveals an additional X-ray component that remains constant during the deep fade and that must therefore arise from a causally disconnected region at this epoch.   

\subsubsection{2020 outburst}

The outburst of April--June 2020 was the second-brightest X-ray outburst of OJ 287 we detected with Swift. During the peak of the outburst, we obtained follow-up spectroscopy with XMM-Newton and NuSTAR and increased the cadence of Swift observations \citep{Komossa2020}.  Our XMM-Newton spectrum firmly established the presence of the super-soft synchrotron component seen in the Swift snapshot spectra and allowed detailed spectral modelling. Our NuSTAR spectrum revealed a spectral component extending to $\sim$70 keV with $\Gamma_{\rm x}=2.2$, remarkably soft for that high-energy band, and softer than pure IC emission. 
The rapid flux variability detected with Swift, with a flux doubling within a few days, is faster than the light-crossing time at the last stable orbit of the accretion disk around the primary SMBH. All these observations establish the non-thermal synchrotron nature of this bright outburst that is accompanied by a radio flare as well (Fig. \ref{fig:Swift-EB-2019-21}). Komossa et al. \cite{Komossa2020} concluded that the timing of the outburst is consistent with an after-flare
predicted by the binary SMBH model \citep{Sundelius1997}. 

\begin{figure}
\begin{center}
\includegraphics[clip, trim=1.9cm 4.9cm 1.0cm 3.4cm,  width=9cm]{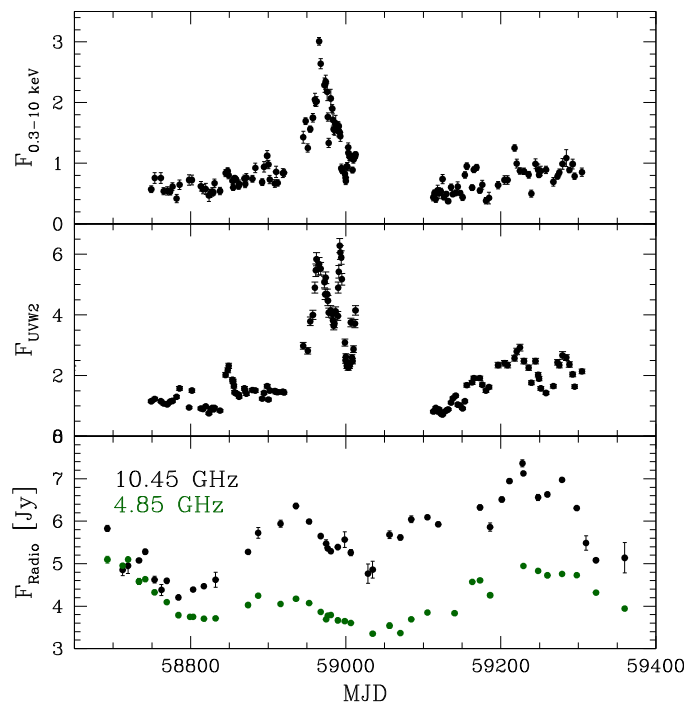}
	\caption{OJ 287 light curve since mid-2019. Upper and middle panel: Swift X-ray and UV(W2) flux in units of 10$^{-11}$ erg s$^{-1}$ cm$^{-2}$.  Lower panel: Effelsberg radio flux density at 10.45 GHz (black) and 4.85 GHz (green).
     }
     \label{fig:Swift-EB-2019-21}
\end{center}
\end{figure}

\subsubsection{Late 2020 -- early 2021 flux evolution and search for precursor flare activity}

Single bright optical flares, referred to as `precursor flares', were seen to precede some of the previous impact flares of OJ 287 \citep{Pihajoki2013}. Since the next of the impact flares is predicted to become visible in 2022 \citep{Dey2018,  Laine2020}, it is interesting to ask whether we detect any precursor flare activity in our most recent Swift light curve. In fact, 
a precursor event was predicted around the epoch 2020.96$\pm$0.10 \citep{Pihajoki2013}. During the time frame of 
2020 October -- 2021 February we detect some of the mini-flaring activity similar in amplitude and duration to previous epochs and none of them stands out. In addition, the light curve also reveals a slow, broad, systematic rise in the UV--optical flux (Figs. \ref{fig:lc-Swift-fluxes2016} and \ref{fig:Swift-EB-2019-21}). The rise started in September 2020 and reached its first maximum in early January 2021. 
While the timing 
agrees with the predicted precursor flare, the observed duration is longer than the sharp events discussed by ref. \citep{Pihajoki2013}. 
The Effelsberg light curve at selected frequencies is shown in Fig. \ref{fig:Swift-EB-2019-21} in comparison with the recent Swift light curve. It demonstrates that the long-lasting optical flare is accompanied by flaring activity in the radio band. Precursor flares could be caused by temporary accretion and jet launching events by the secondary SMBH as it enters the denser parts of the disk corona before it impacts the disk itself \citep{Pihajoki2013, Dey2021}. However, the particular flare we observe is most likely related to a jet component or the core of the primary SMBH, because its radio properties are very similar to the earlier 2020 flare and it would require finetuning to get a near-identical event from a temporary jet of the secondary SMBH.  The cadence of our Swift observations was only 3--4 d during the epoch in question, so we may have either missed any sharp event, or else it may have remained faint (for instance due to a different angle of approach) and was therefore undetectable. 







\section{Summary and Conclusions}

We have presented a description of the project MOMO that aims at understanding  blazar physics and binary  black  hole  physics  of  OJ  287  during  its  recent evolution.
Swift and the Effelsberg radio telescope play a central role in this project and we have used both to obtain measurements of OJ 287 at $>$13 frequencies, along with deeper follow-up spectroscopy from the optical to X-rays in outburst or deep minima states. The rich data sets and their interpretation are presented in a sequence of publications. 
%
Previous work focused on:  the particularly bright 2016/17 Swift outburst \citep[paper I,II,][]{Komossa2017, Komossa2020}; polarimetry \citep[paper Ib,][]{Myserlis2018}; the bright, super-soft, non-thermal 2020 outburst -- interpreted as possible binary after-flare -- with its exceptional spectral components measured with Swift, XMM-Newton and NuSTAR
\citep[paper II,][]{Komossa2020}; two decades of XMM-Newton spectroscopy 
establishing OJ 287 as one of the most spectrally variable blazars in the X-ray band \citep[paper III,][]{Komossa2021a}; and a characterization of the major optical--X-ray variability properties and cross-band lags based on two decades of Swift monitoring \citep[paper IV,][]{Komossa2021c}.
Here, we have further presented our radio observations since 2019 revealing a high level of activity since mid-2019 and our latest Swift multiwavelength results. 
Our observations continue as OJ 287 nears its next impact flare predicted by the binary SMBH model. 

A particular motivation for initiating MOMO has been with the upcoming GW interferometer LISA{\footnote{\url{https://www.elisascience.org/}}} in mind. LISA will directly detect GWs from merging supermassive binaries. As one of the best candidates to date for hosting such a binary, already measurably shrinking due to the emission of GWs, OJ 287 serves as a test bed for scrutinizing and establishing the physics that drive the multi-wavelength emission and orbital evolution of such systems. Independent of any binary's presence, OJ 287 is a nearby bright blazar emitting across the electromagnetic spectrum from the radio to the very-high energy regime, and therefore MOMO -- the densest monitoring of OJ 287 so far carried out at multiple radio, UV and X-ray frequencies -- provides us with new constrains on the disk--jet physics of blazars.


\vspace{6pt} 




\funding{This research received no external funding.} 

\dataavailability{Reduced data are available upon reasonable request. Raw data can be retrieved from the Swift and XMM-Newton archives at \url{https://swift.gsfc.nasa.gov/archive/} and \url{https://www.cosmos.esa.int/web/xmm-newton}, respectively.} 

\acknowledgments{We would like to thank the Swift, Effelsberg, XMM-Newton and NuSTAR teams for approving and carrying out our observations.
SK would like to thank Ski Antonucci, Zoltan Haiman, and Pauli Pihajoki for very useful discussions.  
We acknowledge the use of data we have obtained with the Neil Gehrels Swift mission. We also acknowledge the use of public data from the Swift data archive. 
This work made use of data supplied by the UK Swift Science Data Centre at the University of Leicester \citep{Evans2007}.
This research is partly based on observations obtained with XMM-Newton, an ESA science mission with instruments and contributions directly funded by ESA Member States and NASA. This research is partly based on observations obtained with NuSTAR. 
This work is partly based on data obtained with the 100-m telescope of the Max-Planck-Institut
f\"ur Radioastronomie at Effelsberg.
This research has made use of the
 XRT Data Analysis Software (XRTDAS) developed under the responsibility
of the ASI Space Science Data Center (SSDC), Italy.}

\conflictsofinterest{The authors declare no conflict of interest.} 



\abbreviations{Abbreviations}{
The following abbreviations are used in this manuscript:\\

\noindent 
\begin{tabular}{@{}ll}
BLR & broad-line region \\ 
DCF & discrete correlation function \\ 
EC & external Comptonization\\
EHT &  Event Horizon Telescope \\ 
EUV & extreme ultraviolet \\ 
GR & general relativity \\ 
GWs & gravitational waves \\ 
IC & inverse Compton\\
LBL & low-frequency peaked blazar \\ 
LISA & laser interferometer space antenna \\ 
MOMO & Multiwavelength Observations and Modelling of OJ 287 \\
MJD & modified Julian date \\
PI & principal investigator \\ 
SED & spectral energy distribution \\
SKA & square-kilometer array \\ 
SMBH & supermassive black hole \\
SMBBH & supermassive binary black hole \\
SSC & synchrotron-self-Compton\\
UVOT & UV-optical telescope \\
VLBI & very long baseline interferometry \\ 
XRT & X-ray telescope 
\end{tabular}}

\appendixtitles{no} 

}
\end{paracol}
\reftitle{References}

\end{document}